\documentclass[letter]{aa} 

\usepackage{graphicx}
\usepackage{txfonts}
\usepackage{placeins}
\begin{document}

    \title{Disentangling the origin of chemical differences using GHOST}
   
   \titlerunning{A chemical difference in a giant-giant pair}
   \authorrunning{Saffe et al.}

   \author{C. Saffe\inst{1,2,9}, P. Miquelarena\inst{1,2,9}, J. Alacoria\inst{1,9}, 
           E. Martioli\inst{7}, M. Flores\inst{1,2,9}, M. Jaque Arancibia\inst{3,4}, R. Angeloni\inst{5},
           E. Jofr\'e\inst{6,9}, J. Yana Galarza\inst{8}, E. Gonz\'alez\inst{2} \and A. Collado\inst{1,2,9}
           }

\institute{Instituto de Ciencias Astron\'omicas, de la Tierra y del Espacio (ICATE-CONICET), C.C 467, 5400, San Juan, Argentina
         \and Universidad Nacional de San Juan (UNSJ), Facultad de Ciencias Exactas, F\'isicas y Naturales (FCEFN), San Juan, Argentina
         \and{Instituto de Investigaci\'on Multidisciplinar en Ciencia y Tecnolog\'ia, Universidad de La Serena, Ra\'ul Bitr\'an 1305, La Serena, Chile}
         \and Departamento de F\'isica y Astronom\'ia, Universidad de La Serena, Av. Cisternas 1200 N, La Serena, Chile
         \and Gemini Observatory / NSF’s NOIRLab, Casilla 603, La Serena, Chile
         \and Observatorio Astron\'omico de C\'ordoba (OAC), Laprida 854, X5000BGR, C\'ordoba, Argentina
         \and Laborat\'orio Nacional de Astrof\'isica (LNA/MCTI), rua Estados Unidos 154, Itajub\'a, MG, Brasil
         \and The Observatories of the Carnegie Institution for Science, 813 Santa Barbara Street, Pasadena, CA 91101, USA
         \and Consejo Nacional de Investigaciones Cient\'ificas y T\'ecnicas (CONICET), Argentina
         }

   \date{Received xx, 2024; accepted xx, 2024}

 
  \abstract
   {}
   {We explore different scenarios to explain the chemical difference found in the remarkable giant-giant binary system HD 138202 + CD-30 12303.
   For the first time, we suggest how to distinguish these scenarios by taking advantage of the extensive convective envelopes of giant stars.
   }
   {We carried out a high-precision determination of stellar parameters and abundances by applying a full line-by-line differential analysis
   on GHOST high-resolution spectra.
   We used the FUNDPAR program with ATLAS12 model atmospheres and specific opacities calculated for an arbitrary composition
   through a doubly iterated method. Physical parameters were estimated with the \texttt{isochrones} package and evolutionary tracks
   were calculated via MIST models.}
   {We found a significant chemical difference between the two stars ($\Delta$[Fe/H] $\sim$ 0.08 dex),
which is largely unexpected considering the insensitivity of giant stars to planetary ingestion and diffusion effects.
We tested the possibility of engulfment events by using several different combinations
of stellar mass, ingested mass, metallicity of the engulfed object and different convective envelopes.
However, the planetary ingestion scenario does not seem to explain the observed differences.
For the first time, we distinguished the source
of chemical differences using a giant-giant binary system.
By ruling out other possible scenarios such as planet formation and evolutionary effects between the two stars,
we suggest that primordial inhomogeneities might explain the observed differences.
This remarkable result implies that the metallicity differences that were observed in at least some
main-sequence binary systems might be related to primordial inhomogeneities rather than engulfment events.
We also discuss the important implications of finding primordial inhomogeneities,
which affect chemical tagging and other fields such as planet formation.
We strongly encourage the use of giant-giant pairs. 
They are a relevant complement to main-sequence pairs
for determining the origin of the observed chemical differences in multiple systems.}
   {}

   \keywords{stars: chemically peculiar -- stars: early type -- stars: abundances -- planetary systems 
               }

   \maketitle
%

\section{Introduction}

Most binary systems are thought to have formed from a common molecular cloud
that shared the same chemical composition.
This is supported by observations of binary systems in star-forming regions
\citep[e.g.][]{reipurth07,vogt12,king12} and by numerical models of binary formation
\citep[e.g.][]{kratter11,reipurth-mikkola12}.
However, a number of works studied their chemical composition and found
slight but noticeable differences between the components of some binary systems
\citep[e.g.][]{gratton01,desidera04,desidera06}.

It is difficult to explain the origin of the chemical differences, and a number of possible explanations emerge.
For instance, \citet{melendez09} showed that the Sun is deficient in refractory elements
when compared to solar twins. They suggested that the missing refractories
might be locked up in terrestrial planets.
They noted that the solar depletion pattern might be explained by removing the combined mass of the terrestrial planets from the convective zone alone.
Some studies followed this idea and tried to explain the differences observed in 
main-sequence binary systems where only one component hosts a known planet
\citep[e.g.][]{ramirez11,saffe15} or a debris disk \citep{saffe16}.
Alternatively, \citet{booth-owen20} suggested that a giant planet can block a mass of dust
exterior to its orbit at the time of planet formation, preventing the dust from accreting onto the star.
In this way, these two scenarios \citep{melendez09,booth-owen20}
attributed the chemical differences observed in binary systems to the planet formation process.

Planetary engulfment events also appear to be a promising scenario to explain the observed
differences in main-sequence binary systems \citep[see, e.g.][]{gratton01,saffe17,oh18,nagar20,spina21,jofre21,flores24}.
We note that chemical inhomogeneities detected in the open clusters M67, Hyades, and Pleiades were also ascribed to
ingestion events \citep{oh17,ness18,spina18}.
In particular, \citet{spina21} performed a statistical study of 107 main-sequence binary systems
and suggested that planet engulfment events are common, with an occurrence $\geq$ 20-35\% in solar-like stars.
In stark contrast, \citet{behmard23} claimed that engulfment signatures are rarely detected (a rate closer to $\sim$2.9\%)
and ruled out practically all previous engulfment detections in 10 different binary systems.
Instead of ingestion events, \citet{behmard23} suggested that the differences observed in binary systems may be attributed
to primordial differences between stellar components.
We also note that on more theoretical grounds, chemical signatures of planet ingestion seem unlikely
\citep{theado-vauclair12} because the ingested material is rapidly diluted by thermohaline mixing.
Thus, it is not entirely clear that engulfment events can explain the chemical differences observed
in binary systems.

Primordial inhomogeneities were also suggested based on a study of binary systems 
by \citet{ramirez19} and \citet{liu21}.
The two studies found that binaries with a higher projected separation $d$ present a higher
metallicity difference $|\Delta$[Fe/H]$|$,
and they attributed this correlation to a primordial chemical inhomogeneity.
However, \citet{ramirez19} cautioned that engulfment events could blur the proposed correlation,
while \citet{liu21} noted that atmospheric diffusion could also add a possible dependence on the stellar
parameters for main-sequence stars.
We also note that diffusion effects could produce chemical differences between stars in open clusters \citep[e.g. ][]{liu19,souto19,casamiquela20}.
The studies above illustrate that the origin of the chemical differences
observed in binary systems is strongly debated.
In particular, it is hard to distinguish possible superposed effects 
such as primordial inhomogeneities, planetary ingestion, and diffusion effects
\citep[e.g.][]{liu16,ramirez19,liu21,behmard23,nissen-gustafsson18}.

Most previous works focused on main-sequence binary systems.
However, we note that giant stars are also widely used to perform detailed comparisons in the chemical tagging \citep[e.g. ][]{bovy16,ness18,price-jones19}.
Giant stars could present evolutionary effects in species such as C and N.
Some authors preferred to avoid these particular species in the chemical tagging \citep[e.g. ][]{bovy16}
while others preferred to include them with a caution \citep[e.g. ][]{ness18,price-jones19}.
In any case, it is important to note that giant stars are significantly less prone to diffusion effects than main-sequence stars because their convection 
zones are deeper and more massive \citep[e.g. ][ and references therein]{korn07,dotter17}.
 \citet{dotter17} suggested the use of giant stars (e.g., in the APOGEE survey) to minimize diffusion
effects in the chemical tagging. 
In addition, it is expected that signatures of external pollution
(including their related condensation temperature T$_{c}$ trends) decrease significantly in giants stars
compared to main-sequence stars \citep[e.g. ][]{fischer-valenti05,pasquini07b,spina21}.
In other words, giant stars are thought to be significantly less sensitive than main-sequence stars
to diffusion and pollution effects.
This makes giant-giant pairs ideal targets for testing the chemical homogeneity assumed by the chemical tagging to determine the origin of the observed differences.
For example, it would be highly desirable to explore the correlation suggested
between $d$ and $|\Delta$[Fe/H]$|$ \citep{ramirez19,liu21} using giant-giant pairs,
which would avoid the suggested blur of the correlation due to diffusion and pollution effects.
However, very few works have studied the composition of giant-giant pairs in detail so far \citep[e.g.][]{torres15}. 

We present in this work the first detailed analysis of the chemical composition of the giant-giant binary
HD 138202 + CD-30 12303 (hereafter, stars A and B), with a bound probability $>$95\% from Gaia \textit{EDR3} data \citep{el-badry21}
and a projected separation of 73" ($\sim$38575 au).
Surprisingly, we found a significant chemical difference between the two stars ($\Delta$[Fe/H] $\sim$ 0.08 dex),
which is largely unexpected considering their insensitivity to planetary ingestion and diffusion effects.
We consider this a remarkable result for several reasons.
First, we found a difference between the two components in a binary system in which we separated the source of the chemical differences for the first time
(and we attributed them to primordial inhomogeneities).
This would imply that metallicity differences should in general 
be taken with caution because they might originate in scenarios that are unrelated to engulfments.
Thus, in order to claim ingestion events, it would become mandatory to explore 
additional evidence such as the study of T$_{c}$ trends, the lithium content, and the stellar rotation.
Second, if it is confirmed that the slight chemical differences we detected are primordial
\citep[which would be more in line with the results of ][]{behmard23},
it could have important implications as follows.
It would challenge the main assumption of chemical tagging, that is, the chemical homogeneity. 
It would place important constraints on formation models of multiple systems \citep[e.g. ][]{bate19,guszejnov21} and 
interstellar medium (ISM) mixing \citep[e.g. ][]{feng-krumholz14,armillotta18}.
Finally, a primordial difference could severely impact planet formation,
which might help to explain, for example, why very similar stars in wide binary systems present 
different planetary systems \citep[e.g. ][]{biazzo15,teske16}.
This highlights that the study of giant-giant pairs (as in the present work) may emerge as an important complement to the study of main-sequence pairs.



\section{Observations and data reduction}

Observations of the giant-giant pair were acquired through the
Gemini High-resolution Optical SpecTrograph (GHOST),
which is attached to the 8.1 m Gemini South telescope at Cerro Pach\'on, Chile.
GHOST is illuminated via 1.2" integral field units that provide the
input light apertures. The spectral coverage of GHOST between
360-900 nm is appropriate for deriving stellar parameters and
chemical abundances using several features. It provides a high resolving power
R$\sim$50000 in the standard resolution 
mode\footnote{https://www.gemini.edu/instrumentation/ghost}.
The read mode was set to medium, as recommended for relatively bright
targets.
The observations were taken on May 15, 2023, during a GHOST science verification run,
using the same spectrograph configuration for both stars.
The exposure times were 3 $\times$ 200 sec and 3 $\times$ 180 sec
on targets A and B, obtaining a final signal-to-noise ratio 
(S/N) of $\sim$400 per pixel measured at $\sim$6000 {\AA} in the combined spectra for both stars.
The standard star 18 Sco was also observed with the same spectrograph setup, 
achieving a similar S/N to use as initial reference.
The spectra were reduced using the GHOST data reduction pipeline v1.0.0, which works under 
DRAGONS\footnote{https://www.gemini.edu/observing/phase-iii/reducing-data/dragons-data-reduction-software}. This is
a platform for the reduction and processing of astronomical data.

\section{Stellar parameters and abundance analysis}

\begin{figure}
\centering
\vskip -0.2in
\includegraphics[width=6cm]{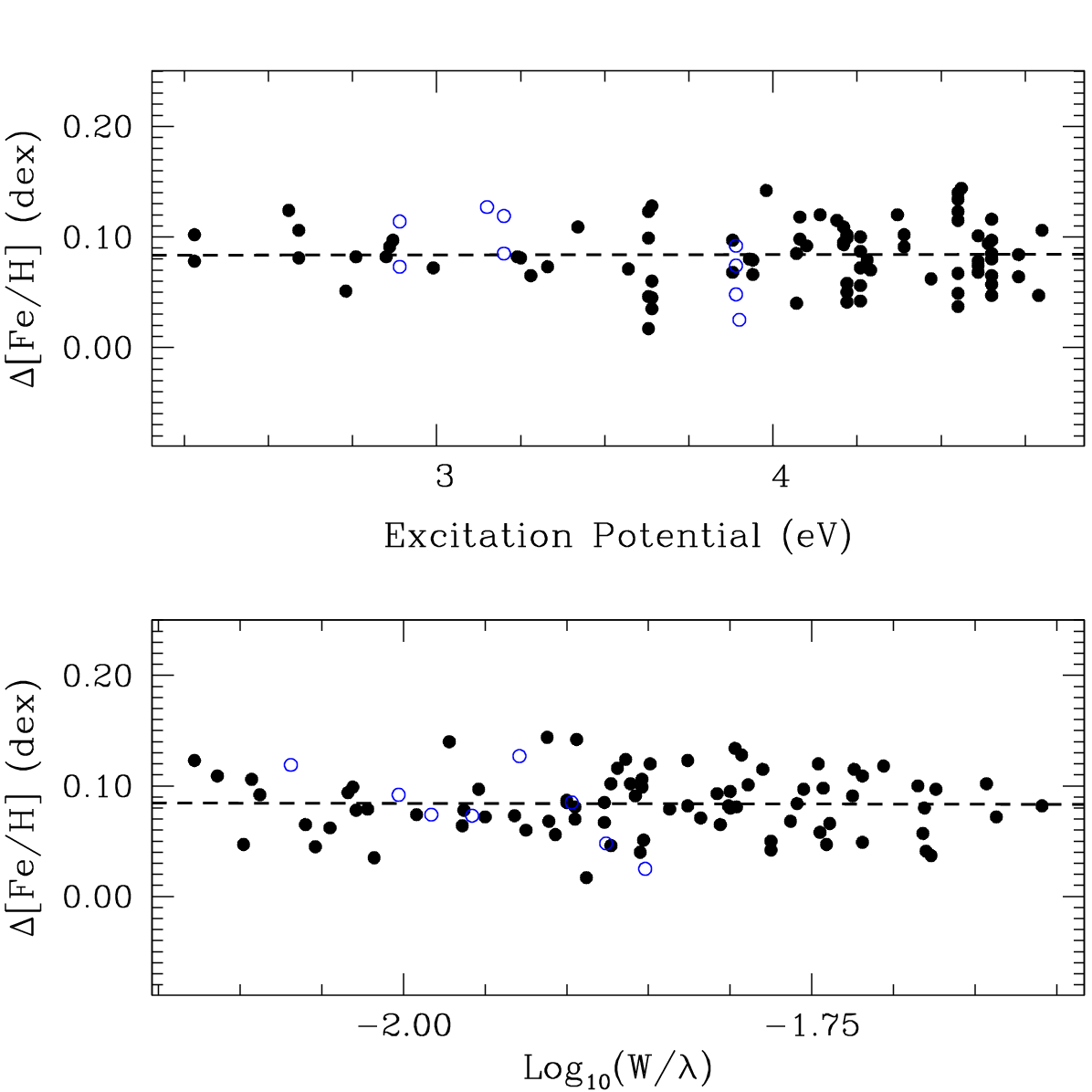}
\caption{Differential abundance vs. excitation potential (upper panel) 
and vs. reduced EW (lower panel) for the case (A - B).
The filled and empty circles correspond to \ion{Fe}{I} and \ion{Fe}{II}.}
\label{equil-relat}%
\end{figure}

The fundamental parameters (T$_{eff}$, {log $g$}, [Fe/H], and v$_{turb}$)
were derived following a similar procedure as in our previous work \citep{saffe18,saffe19}.
We measured the equivalent widths (EW) of the metallic lines 
using the IRAF task \texttt{splot} in the stellar spectra.
The line lists were taken from works of giant stars \citep{jofre15,soto21} with updated laboratory data
for some lines \citep{liu14,melendez14,bedell14}.
We imposed an excitation and ionization balance of \ion{Fe}{} lines using the differential
version of the FUNDPAR program \citep{saffe11,saffe18}.
This code made use of ATLAS12 model atmospheres \citep{kurucz93} together with the MOOG program \citep{sneden73},
using specific opacities calculated for an arbitrary composition.
The stellar parameters were determined using the solar twin 18 Sco as reference, that is, (A - 18 Sco) and (B - 18 Sco),
by adopting (5811 K, 4.42 dex, 0.04 dex, 1.06 km s$^{-1}$) for 18 Sco. These are the highest-precision parameters currently derived
for this solar twin \citep{liu20}.
We recalculated the parameters of star A, but used B as reference, that is, (A - B), 
as listed in Table \ref{stellar.params}\footnote{$\Delta$(Fe/H) $=$ log(Fe/H)$_{*}$ $-$ log(Fe/H)$_{ref}$} and as shown in Appendix A.
The errors on the stellar parameters were derived following the procedure detailed in \citet{saffe15},
which takes the individual and mutual covariance terms of the error propagation into account.
Figure \ref{equil-relat} shows the abundance versus excitation potential (top panel)
and the abundance versus the reduced EW (bottom panel) for the case (A - B)
with an average uncertainty of 0.02 dex.


\begin{table}
\centering
\caption{Stellar parameters derived for each star.}
\vskip -0.1in
\scriptsize
\begin{tabular}{ccccc}
\hline
\hline
 (Star - Reference) & T$_{eff}$ & log $g$ & $\Delta$(Fe/H) & v$_{turb}$ \\
  & [K] & [dex] & [dex] & [km s$^{-1}$] \\
\hline
(A - 18 Sco)   &  4952 $\pm$ 49 & 2.51 $\pm$ 0.09 &  0.067 $\pm$ 0.014 & 1.82 $\pm$ 0.06 \\
(B - 18 Sco)   &  5007 $\pm$ 45 & 2.57 $\pm$ 0.08 & -0.015 $\pm$ 0.012 & 1.72 $\pm$ 0.05 \\
(A - B)        &  4952 $\pm$ 30 & 2.52 $\pm$ 0.06 &  0.083 $\pm$ 0.011 & 1.82 $\pm$ 0.04 \\
\hline
\end{tabular}
\normalsize
\label{stellar.params}
\end{table}

The hyperfine structure splitting (HFS) was considered for 
\ion{V}{I}, \ion{Mn}{I}, \ion{Co}{I}, and \ion{Cu}{I} 
using the HFS constants of \citet{kurucz-bell95} and performing spectral synthesis for these species.
We also derived the \ion{Li}{I} abundance using spectral synthesis with the resonance line 6707.80 {\AA,}
which includes the doublet 6707.76 {\AA}, 6707.91 {\AA} and HFS components.
We corrected the \ion{Li}{I} abundance for nonlocal thermodynamic equilibrium (NLTE) effects by
interpolating in the data of \citet{lind09}.
We adopted $\Delta$Li$_{NLTE-LTE}$ $\sim$ 0.18 dex for both stars, implying that NLTE effects
are not significant for the case (A - B).
The C/N ratio, which is sensitive to evolutionary effects, was derived through spectral synthesis.
The C abundances were obtained from the C I lines between 7111 - 7116 \AA, while N abundances
were obtained by fixing the C content and then varying N in the CN band near $\sim$4212 \AA.


\section{Results and discussion}

\begin{figure}
\centering
\includegraphics[width=7cm]{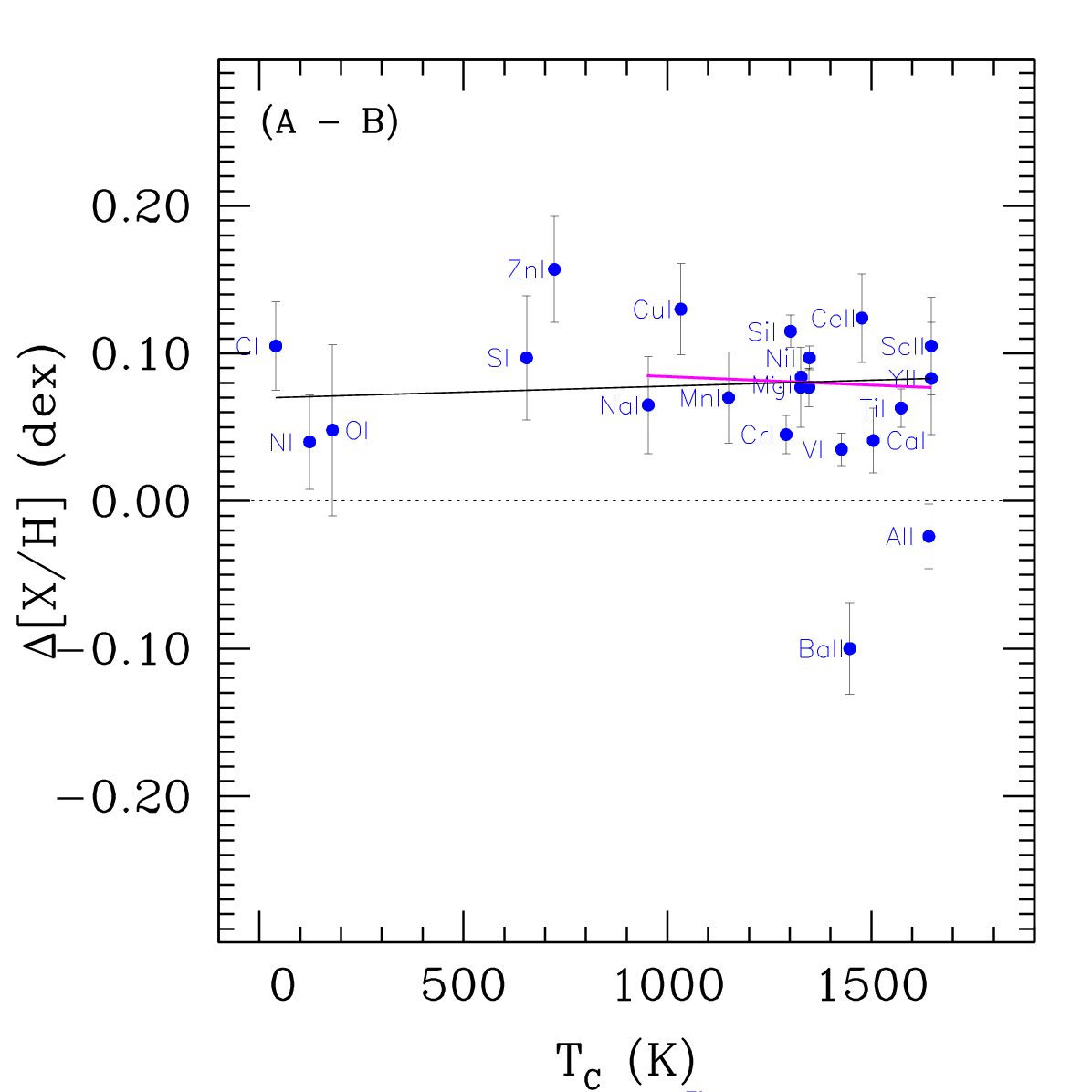}
\caption{Differential abundances (A - B) as a function of T$_{c}$.
The continuous lines show linear fits to all species (black) and to the refractory species (T$_{c}$ $>$ 950 K, magenta).
The dotted black line corresponds to a null metallicity difference between the stars.}
\label{relat.tc}%
\end{figure}

The star A is more metal rich by $\sim$0.08 dex than its companion (see Table \ref{stellar.params}).
The condensation temperatures of the elements were taken from the 50\% T$_{c}$ values derived by \citet{lodders03}
for a solar system gas with [Fe/H]=0. 
The differential abundances (A - B) are presented in Fig. \ref{relat.tc} as a function of T$_{c}$.
We discarded from the fits species with evolutionary effects such as C and N, 
and species that could drive the trends, such as \ion{Zn}, \ion{Al}\ and \ion{Ba}.
The black dotted line would correspond to a null metallicity difference between the stars.
We obtained slopes of 0.80$\pm$6.68 10$^{-5}$ dex K$^{-1}$ and -1.20$\pm$13.20 10$^{-5}$ dex K$^{-1}$
for all species and for the refractory species, respectively.
The abundance dispersion in Fig. \ref{relat.tc} tends to increase the uncertainties
in the slopes, which should be taken with caution.
The average abundances of the volatiles (T$_{c}$ $<$ 950 K) and refractories are {0.073$\pm$0.035} dex and {0.081$\pm$0.010} dex,
respectively. Although we find no significant T$_{c}$ trend between stars A and B,
most species differ by $\sim$0.08 dex in the two stars.
In the next subsections, we explore different scenarios to explain the observed differences.


\subsection{Scenario of the evolutionary state}

If both giant stars underwent first dredge up (FDU) before, 
as is strongly suggested by the low C/N and Li abundances \citep{salaris-cassisi17},
evolutionary effects ($<$0.01 dex) cannot explain the observed difference of $\sim$0.08 dex in Fe.
Even when we assume that only one star passed through the FDU phase (which is highly unlikely),
evolutionary effects ($<\sim$0.04 dex) are not enough to explain the observed difference in Fe.
The calculations are extensively discussed in Appendix B.

\subsection{Scenario of planet ingestion}

A metallicity variation due to a planet engulfment is difficult to detect in stars with a significant convective envelope
\citep[e.g. ][]{fischer04,fischer-valenti05,pasquini07a,pasquini07b,spina15,spina21}.
For example, \citet{fischer04} and \citet{fischer-valenti05} suggested that the higher convective envelopes of subgiant stars
(compared to main-sequence stars) should result in a much lower or null increase in metallicity due to external pollution.
Moreover, \citet{pasquini07a,pasquini07b} estimated that the convective envelope of a main-sequence 1 M$_{\odot}$ star increases
about 30-35 times when the star evolves to the RGB, changing from $\sim$0.022 M$_{\odot}$ to $\sim$0.77 M$_{\odot}$.
The authors claimed that a possible higher metallicity (confined to the superficial layers) would easily
decrease to the primordial values through the deepening convective zone (CZ). 
In particular, they estimated that an excess of 0.25 dex in a solar star would become lower than $\sim$1\% in a giant star.
Similarly, for the case of ingestion in main-sequence stars, \citet{spina15} showed that the mass of the CZ
is a critical parameter for determining the [Fe/H] enhancement (e.g., their Figure 2).
More recently, \citet{spina21} showed that planetary ingestion is extremely sensitive to the thickness of the external layer
(e.g., their Fig. 1), and they concluded that giant stars (and low-mass main-sequence stars) are not expected to vary their
chemical composition because their external layers are extremely thick.
In other words, giant stars are thought to be significantly less sensitive than main-sequence stars to engulfment events.


These arguments disfavor an engulfment event to explain the observed metallicity difference between stars A and B. 
However, we estimated the amount of planetary material that star A would  need to ingest to increase its metallicity by $\sim$0.08 dex compared to star B.
A first-order approximation is detailed in Appendix B.
We estimate that star A would need to have ingested between 11.0$-$150.0 M$_{Jup}$ of planetary material,
depending on the adopted convective envelope mass and metallic content of the ingested planet (see Table \ref{ingested.mass}).
In particular, the ingestion of planets with masses near the lower value of 11 M$_{Jup}$ would require extremely metallic objects,
which are difficult to explain with current planet formation models \citep[see e.g. the discussions in ][]{leconte09,thorngren16}.
In this case, a low convective envelope mass for star A of only 35\% of their total mass would also be required.
In addition, the ingestion of massive planetary bodies should produce a significant increase in the rotational
velocity of giant stars \citep{carlberg09,carlberg12,privitera16,stephan20}.
We estimate an increase in the rotational velocity of star A between 8.72$-$37.88 km s$^{-1}$ after the planet ingestion (see Table \ref{ingested.mass}).
However, we find no evidence of significantly different projected rotational velocities between stars A and B 
\citep[v sini 4.73$\pm$0.44 km s$^{-1}$ and 5.19$\pm$0.34 km s$^{-1}$,
after quadratically subtracting v$_{macro}$ from ][]{hekker-melendez07}.
Star B presents an even slightly higher v sini than star A, in contrast to what is expected after an ingestion by star A
(although we caution that the axis orientation could still play a role).

We also estimated the expected increase in the lithium content of star A
produced by the planet ingestion (see Table \ref{ingested.mass}).
After the engulfment of 11.0$-$150.0 M$_{Jup}$ of planetary material,
we estimate that the lithium abundance of star A should increase by $\sim$1.50 to 2.10 dex.
However, the Li abundances are A(Li)$_{A}=$0.76$\pm$0.04 dex and A(Li)$_{B}=$1.25$\pm$0.04 dex
\citep[estimated with spectral synthesis and NLTE corrected following ][]{lind09}.
In other words, the planet ingestion predicts a strong increase in lithium of star A,
while the observations seem to show the opposite, that is to say, a higher lithium content in star B.
We note that some works suggested that an ingestion event could be ruled out if the same component
of a binary system presented both a higher Li content and a lower metallicity of the pair
\citep{ramirez19,spina21}, as observed in the stars of the present work.
The Li abundances of stars A and B also appear to be similar to those of other giant stars
(see Fig. \ref{li.vs.params.fig} in Appendix B), while the expected Li value
after an ingestion is considerably higher.
The higher Li abundance of star B compared to star A might at least partially be explained by their slightly higher T$_{eff}$ (see Appendix B).
It is also important to note that the lithium abundances are strongly modified by evolutionary effects
such as the FDU and other still uncertain processes.
 \citet{privitera16} suggested using a possible increase in the rotational velocities rather
than the lithium abundances in order to detect a planetary ingestion event in giant stars.

To summarize the ingestion scenario, 
an engulfment that produced the metallicity difference between stars A and B would require
very high values of ingested material together with extremely metallic planets
and a very low convective envelope of star A. This is highly unlikely.
These extreme values should also be accompanied by a significant increase in the rotational velocity 
and lithium abundances, which are not detected.
The iron and lithium abundances also seem to contradict the abundances expected after an ingestion event.
In addition, no planets are detected around stars A and B.
Therefore, current evidence does not support the ingestion scenario.

\subsection{Scenario of sequestered refractories}

\citet{melendez09} found that the Sun is depleted in refractory species
when compared to solar twins, suggesting that the missing refractories
might be locked in terrestrial planets and in the cores of giant planets.
The depletion pattern is explained if the combined mass
of terrestrial planets is removed from the convective zone of the Sun
\citep{melendez09,chambers10,kunitomo18,bitsch18}.
When the star evolves off the main-sequence, the massive convective
zone in the red giant phase erases this chemical fingerprint,
as suggested by \citet{maldonado-villaver16}.
This implies that the mutual chemical difference found between the giant components in this work
(where no clear T$_{c}$ trend or planets are detected)
can hardly be attributed to this effect.

We also note that the massive convective envelopes of giant stars could present an additional advantage 
for testing the original homogeneity of the chemical tagging.
We mentioned that giants stars are less sensitive to diffusion and pollution effects
than main-sequence stars. In addition, if planet formation produces
a possible sequestration of refractories in the most superficial layers of main-sequence stars
\citep[as suggested by e.g. ][]{melendez09,chambers10,kunitomo18,bitsch18}, 
this fingerprint should be diluted by the massive envelopes of 
giant stars.

\subsection{Scenario of primordial inhomogeneities}

Some works suggested that the slight chemical differences found between the components of 
main-sequence binary systems could be attributed to primordial inhomogeneities \citep{ramirez19,liu21,behmard23}.
We suggest that the chemical differences found between stars A and B are also possibly primordial
by previously ruling out other scenarios.
The large separation of the two stars ($\sim$38575 au) also supports this idea.
However, no T$_{c}$ trend is detected between stars A and B.
In this way, the main finding of this work is the detection of clear chemical differences
but null T$_{c}$ trends in a giant-giant pair.

Further relevant implications of this result are listed below. \\
(a) In general, binaries with overall abundance differences but no clear T$_{c}$ trends are likely due to 
primordial inhomogeneities \citep[many cases of chemical anomalies in different literature works, such as e.g. ][]{spina21}.\\
(b) This poses a crucial challenge to the concept of chemical tagging. Primordial inhomogeneties very likely stem
from an inhomogeneous ISM, and therefore, the origin of stars cannot be easily tagged. 
We derived the 3D velocity difference $\Delta$v between stars A and B and
obtained 1.13 km/s, which would support the binarity (i.e., the conatal nature) of this pair.
For example, \citet{yong23} studied a sample of 125 comoving pairs, selected by taking $\Delta$v $<$ 2.0 km/s,
a cutoff chosen to avoid nonconatal pairs.\\
(c) Primordial inhomogeneities would place important constraints on formation models of multiple systems \citep{bate19,guszejnov21}
and on hydrodynamic simulations of ISM mixing \citep[e.g. ][]{feng-krumholz14,armillotta18}.
For example, \citet{bate19} showed that a lower metallicity results in an increased fragmentation in cores,
filaments, and disks because the cooling rate of dense gas is higher.\\
(d) A primordial difference might also affect planet formation. This might
help to explain, for example, why very similar stars in wide binary systems can present 
different planetary systems \citep[e.g. ][]{biazzo15,teske16}.

We strongly encourage the study of giant-giant pairs. This novel approach
might help us to distinguish the origin of the slight chemical differences 
observed in multiple systems.

\begin{acknowledgements}
We thank the anonymous referee for constructive comments that improved the paper.
The authors thank R. Kurucz and C. Sneden for making their codes available to us.
CS acknowledge the financial support from projects PIP 11220210100048CO and CICITCA 21/E1235.
E.M. acknowledges funding from FAPEMIG under project number APQ-02493-22 and a research productivity grant number 309829/2022-4 awarded by the CNPq, Brazil.
PM and JA acknowledge the financial support from CONICET in the forms of Doctoral Fellowships.
GHOST was built by a collaboration between Australian Astronomical Optics at Macquarie University, National Research Council Herzberg of Canada,
and the Australian National University, and funded by the International Gemini partnership. 
The instrument scientist is Dr. Alan McConnachie at NRC, and the instrument team is also led by Dr. Gordon Robertson (at AAO), and Dr. Michael Ireland (at ANU).
The authors would like to acknowledge the contributions of the GHOST instrument build team, the Gemini GHOST instrument team, the full SV team, 
and the rest of the Gemini operations team that were involved in making the SV observations a success.

\end{acknowledgements}

\begin{appendix}

\FloatBarrier
\clearpage

\section{Stellar parameters and abundances}

We present in Table \ref{tab.abunds} the differential abundances [X/H] $\pm$ $\sigma_{TOT}$ obtained for the case (A-B).
For each species, we show the observational error $\sigma_{obs}$
(estimated as $\sigma/\sqrt{(n-1)}$ , where $\sigma$ is the standard deviation of the different lines)
as well as internal errors due to uncertainties in the stellar parameters
$\sigma_{par}$ (by adding quadratically the abundance variation when modifying
the stellar parameters by their uncertainties). For chemical species with only one line,
we adopted for $\sigma$ the average standard deviation of the other elements.
The total error $\sigma_{TOT}$ was obtained by quadratically adding  $\sigma_{obs}$ and $\sigma_{par}$.

\begin{table}
\centering
\caption{Chemical abundances for the case (A-B).}
\scriptsize
\begin{tabular}{lrcc}
\hline
\hline
Specie     & [X/H] $\pm$ e$_{tot}$ & $\sigma_{obs}$ & $\sigma_{par}$ \\
\hline
{[Li I/H]} & -0.490 $\pm$ 0.031 & 0.010 & 0.029 \\
{[C I/H]}  &  0.105 $\pm$ 0.030 & 0.007 & 0.029 \\
{[N I/H]}  &  0.040 $\pm$ 0.032 & 0.010 & 0.029 \\
{[O I/H]}  &  0.048 $\pm$ 0.058 & 0.010 & 0.057 \\
{[Na I/H]} &  0.065 $\pm$ 0.033 & 0.021 & 0.026 \\
{[Mg I/H]} &  0.077 $\pm$ 0.027 & 0.010 & 0.025 \\
{[Al I/H]} & -0.024 $\pm$ 0.022 & 0.010 & 0.019 \\
{[Si I/H]} &  0.115 $\pm$ 0.011 & 0.010 & 0.005 \\
{[S I/H]}  &  0.097 $\pm$ 0.042 & 0.010 & 0.041 \\
{[Ca I/H]} &  0.041 $\pm$ 0.022 & 0.012 & 0.018 \\
{[Sc II/H]} & 0.105 $\pm$ 0.033 & 0.025 & 0.022 \\
{[Ti I/H]} &  0.063 $\pm$ 0.013 & 0.008 & 0.010 \\
{[V I/H]}  &  0.035 $\pm$ 0.011 & 0.006 & 0.009 \\
{[Cr I/H]} &  0.045 $\pm$ 0.013 & 0.008 & 0.010 \\
{[Mn I/H]} &  0.070 $\pm$ 0.031 & 0.010 & 0.029 \\
{[Fe I/H]} &  0.083 $\pm$ 0.004 & 0.003 & 0.003 \\
{[Fe II/H]} & 0.083 $\pm$ 0.020 & 0.012 & 0.016 \\
{[Co I/H]} &  0.077 $\pm$ 0.013 & 0.007 & 0.010 \\
{[Ni I/H]} &  0.097 $\pm$ 0.008 & 0.006 & 0.004 \\
{[Cu I/H]} &  0.130 $\pm$ 0.031 & 0.010 & 0.029 \\
{[Zn I/H]} &  0.157 $\pm$ 0.036 & 0.010 & 0.034 \\
{[Y II/H]} &  0.083 $\pm$ 0.038 & 0.010 & 0.036 \\
{[Ba II/H]}& -0.100 $\pm$ 0.031 & 0.010 & 0.029 \\
{[Ce II/H]}&  0.124 $\pm$ 0.030 & 0.010 & 0.029 \\
\hline
\end{tabular}
\normalsize
\label{tab.abunds}
\end{table}

We also took the opportunity and estimated T$_{eff}$ of stars A and B by using different photometric calibrations
in order to compare them with high-precision spectroscopic results.
\citet{ramirez-melendez05} provided the T$_{eff}$ calibration as a function of (B-V) through a polynomial fit.
We derived an extinction of A$_{v}\sim$ 0.57 mag from \citet{gaiaDR3} and A$_{v}\sim$ 0.73 mag from the maps of \citet{schlegel98}.
By using A$_{v}$ from Gaia, we obtain T$_{eff}$(A)$=$ 4944$\pm$126 K and T$_{eff}$(B)$=$ 4991$\pm$128 K,
while with A$_{v}$ from \citet{schlegel98}, we obtain slighly higher values, T$_{eff}$(A)$=$ 5054$\pm$132 K and T$_{eff}$(B)$=$ 5104$\pm$140 K.
We note that the spectroscopic T$_{eff}$ agrees well compared to the photometric T$_{eff}$ with A$_{v}$ taken from Gaia. 
This indicates that the extinction values from Gaia might be more appropriate.
By using this photometric estimation of T$_{eff}$, the difference between the two stars would be
T$_{eff}$(B) $-$ T$_{eff}$(A) $\sim$ 47 $\pm$ 179 K, which agrees well with the spectroscopic difference of 55 $\pm$ 66 K.
However, the photometric estimation of the stellar parameters should be taken with caution due to the significant extinction in the direction of the stars.
Stars A and B have a distance of 523$\pm$6 pc and 536$\pm$6 pc, respectively, based on Gaia \textit{EDR3} parallaxes.
These distant stars could lead to differences between spectroscopic and photometric parameters
and should be taken with caution.

We note that star B is enriched in Al and Ba compared to star A, as we shown in Fig. \ref{relat.tc}.
It would be tempting to assume that they were contributed from an unknown AGB companion around star B.
However, it is difficult to precisely estimate the origin of these elements.
Ba is mostly considered an s-process element, that is, produced by a slow neutron-capture reaction of relatively heavy nuclei
and carried to the ISM by the winds of AGB stars.
The isotope $^{26}$Al could also be produced by AGB stars, but other possible
sources are winds of massive and very massive stars and supernova explosions \citep[see e.g.][ and references therein]{martinet22}.
In principle, a common origin for the two elements therefore cannot be entirely discarded.
However, this hypothetical scenario should be taken with caution because no AGB companion
is currently detected orbiting star B.

\section{Scenario of the evolutionary state}

\begin{figure}
\centering
\includegraphics[width=8cm]{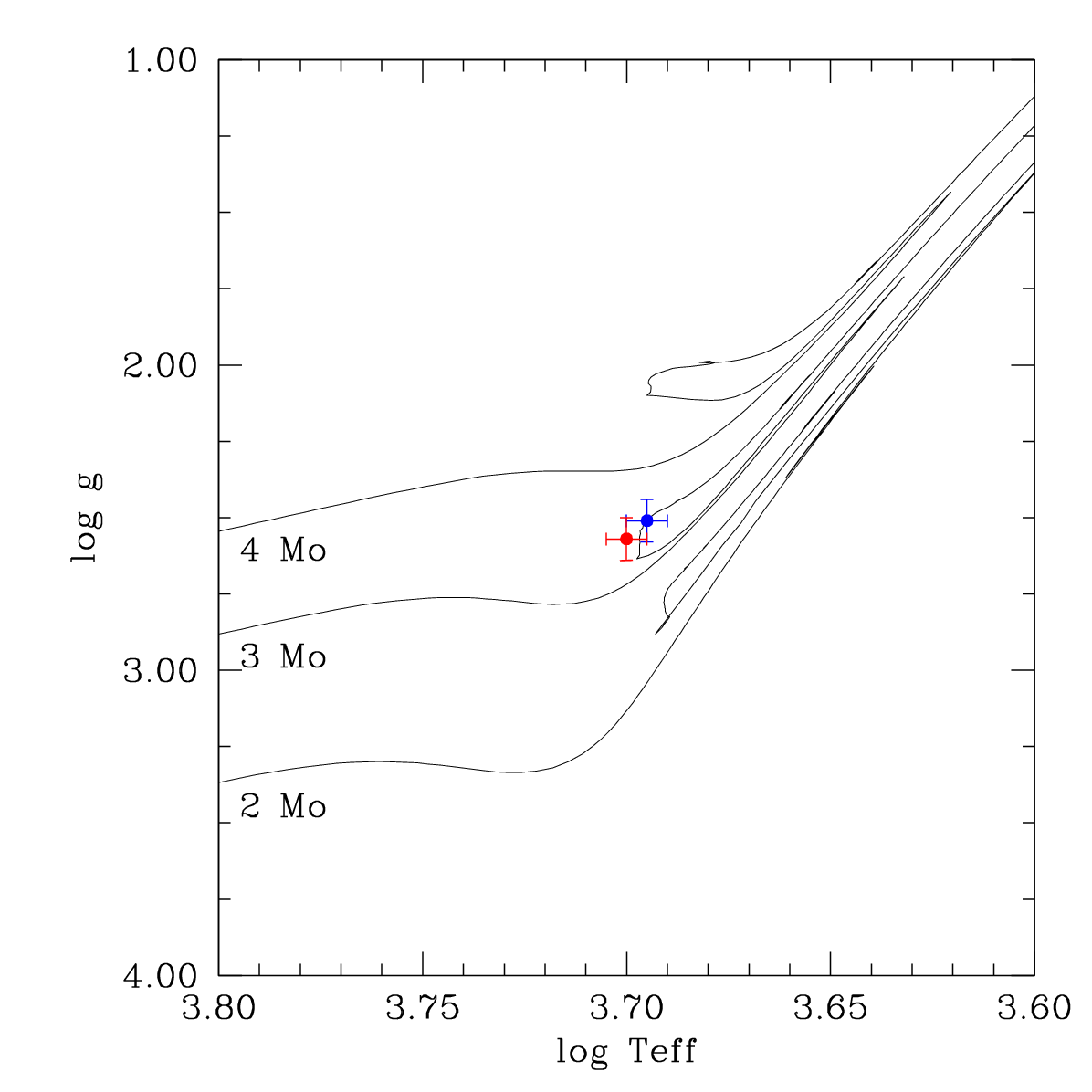}
\caption{log T$_{eff}$ $-$ log g diagram for stars A and B (blue and red, respectively),
and three (solar metallicity) evolutionary tracks corresponding to 2 M$_{\odot}$, 3 M$_{\odot}$, and 4 M$_{\odot}$
calculated using MIST.}
\label{Tracks.234.fig}%
\end{figure}

We explore in this section whether evolutionary effects might explain the observed chemical differences between stars A and B.
RGB stars burn hydrogen in a shell around a inert helium core \citep{iben68},
while red clump (RC) stars are in the stage of core-helium burning \citep{cassisi-salaris97,girardi98}.
RGB and RC stars overlap significantly in the T$_{eff}$ $-$ log g diagram \citep[e.g. ][]{girardi16}
and it is hard to distinguish the two types of stars.
Figure \ref{Tracks.234.fig} shows the log T$_{eff}$ $-$ log g diagram for stars A and B
(blue and red, respectively)
and three (solar metallicity) evolutionary tracks corresponding to 2 M$_{\odot}$, 3 M$_{\odot}$, and 4 M$_{\odot}$,
calculated using the MESA isochrones and stellar tracks 
\citep[MIST, ][]{dotter16}\footnote{https://waps.cfa.harvard.edu/MIST/} version 1.2.
In Fig. \ref{Tracks.234.fig}, stars A and B are located between the tracks of 2 M$_{\odot}$ and 4 M$_{\odot}$.
However, it is difficult to determine a precise value of the mass (and evolutionary state),
considering the mentioned overlap of RGB and RC stars in the T$_{eff}$ $-$ log g diagram \citep[e.g. ][]{girardi16}.

\begin{figure*}
\centering
\includegraphics[width=8cm]{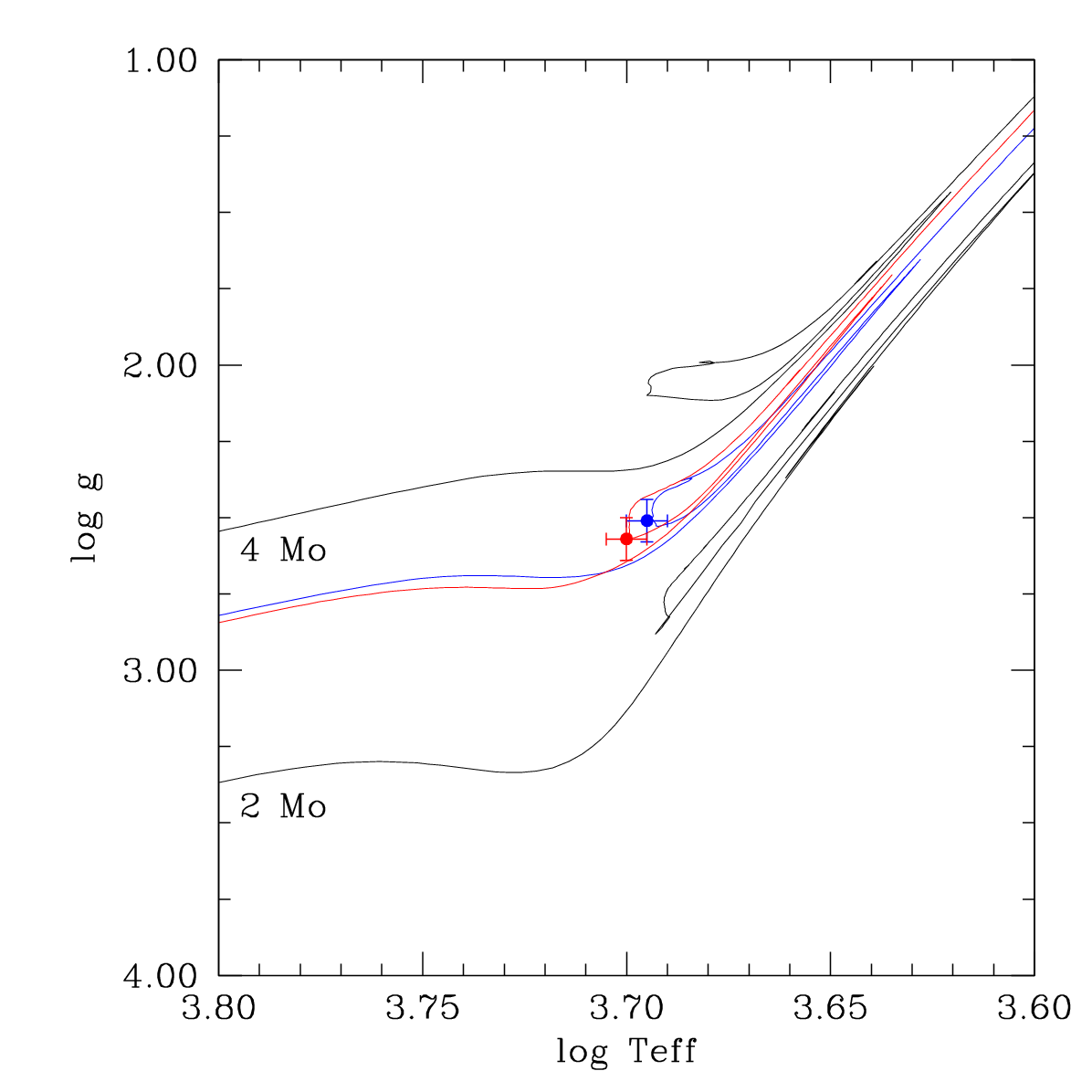}
\includegraphics[width=8cm]{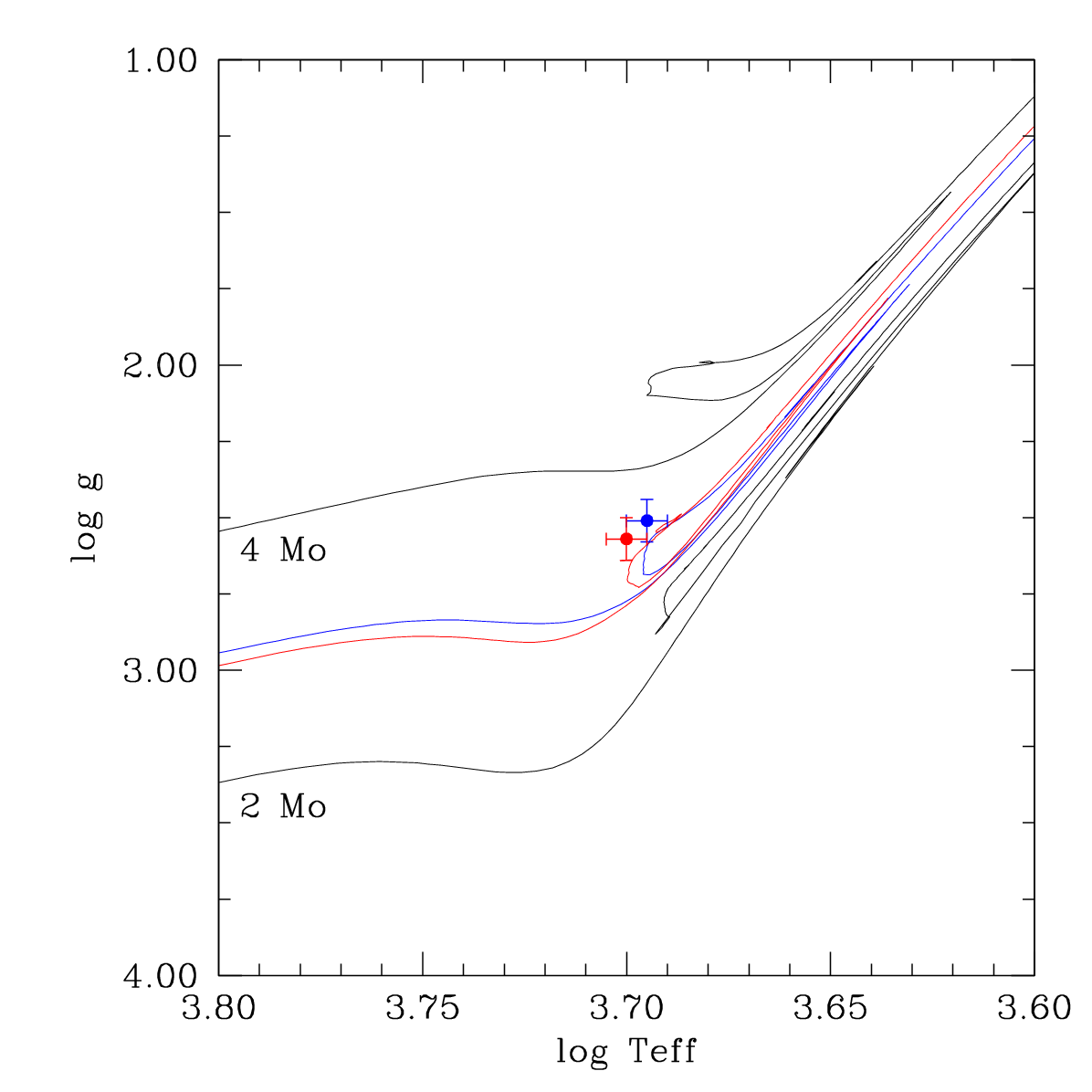}
\caption{log T$_{eff}$ $-$ log g diagram for stars A and B, and the evolutionary tracks
calculated with MIST for their corresponding masses and metallicities. Blue and red correspond to
stars A and B. The plot also shows two additional tracks for 2 M$_{\odot}$ and 4 M$_{\odot}$ (black lines) for reference.
The left and right panels correspond to the RGB and RC cases, respectively.}
\label{Tracks.RGB.RC.fig}%
\end{figure*}

Physical parameters such as mass, age, and evolutionary state were estimated 
using the latest version of the \texttt{isochrones}\footnote{https://github.com/timothydmorton/isochrones} package \citep{morton15}.
The package creates a model for each star and interpolates it in a grid of MIST tracks, indicating its evolutionary state
through equivalent evolutionary points (EEPs), as described in \citet{dotter16}.
In this scheme, the terminal-age main sequence (TAMS), the tip of the RGB (RGBTip), the zero-age core-helium burning (ZACHeB),
and the terminal-age core-helium burning (TACHeB) correspond to EEP values of 454, 605, 631, and 707, respectively.
We used as input for \texttt{isochrones} the high-precision stellar parameters (T$_{eff}$, log g, and [Fe/H]), 
together with the V magnitudes (9.71$\pm$0.03 and 9.79$\pm$0.02 for stars A and B) and the reddening A$_{V}$. 
We note that the extinction in the direction of the stars is significant: 
We obtained A$_{V}$ $\sim$ 0.57 \citep{gaiaDR3} or A$_{V}$ $\sim$ 0.73 \citep{schlegel98}.
However, the physical parameters we derived (mass, age, and evolutionary state) remain almost unchanged 
when the high-precision spectroscopic values alone are used (T$_{eff}$, log g, and [Fe/H]).
We present in Table \ref{mass.age.tab} the mass and age of stars A and B, estimated using \texttt{isochrones},
considering the stars in the RGB or RC phases.
Depending on whether the stars belong to the RGB or RC phases, their physical parameters are different
and should be taken with caution.
In Fig. \ref{Tracks.RGB.RC.fig} we present the log T$_{eff}$ $-$ log g diagram for stars A and B and the evolutionary tracks
calculated with MIST for their corresponding masses and metallicities. Blue and red correspond to
stars A and B. The plots also show two additional tracks for 2 M$_{\odot}$ and 4 M$_{\odot}$ (black lines) for reference.
The left and right panels correspond to the RGB and RC cases, respectively.

\begin{table}
\centering
\caption{Physical parameters estimated with the \texttt{isochrones} package.}
\begin{tabular}{lccccc}
\hline
\hline
Star &     RGB      &     RGB  &     RC       &     RC   \\
     &    Mass      &     Age  &    Mass      &    Age   \\
     &[M$_{\odot}$] &  [Myr]   &[M$_{\odot}$] &  [Myr]   \\
\hline
A    & 3.18$_{-0.36}^{+0.36}$ & 327$_{-87}^{+118}$  & 2.87$_{-0.29}^{+0.29}$ & 488$_{-134}^{+184}$ \\
B    & 3.06$_{-0.27}^{+0.27}$ & 347$_{-73}^{+93}$   & 2.72$_{-0.24}^{+0.24}$ & 540$_{-133}^{+177}$ \\
\hline
\end{tabular}
\label{mass.age.tab}
\end{table}

\begin{figure}
\centering
\includegraphics[width=8cm]{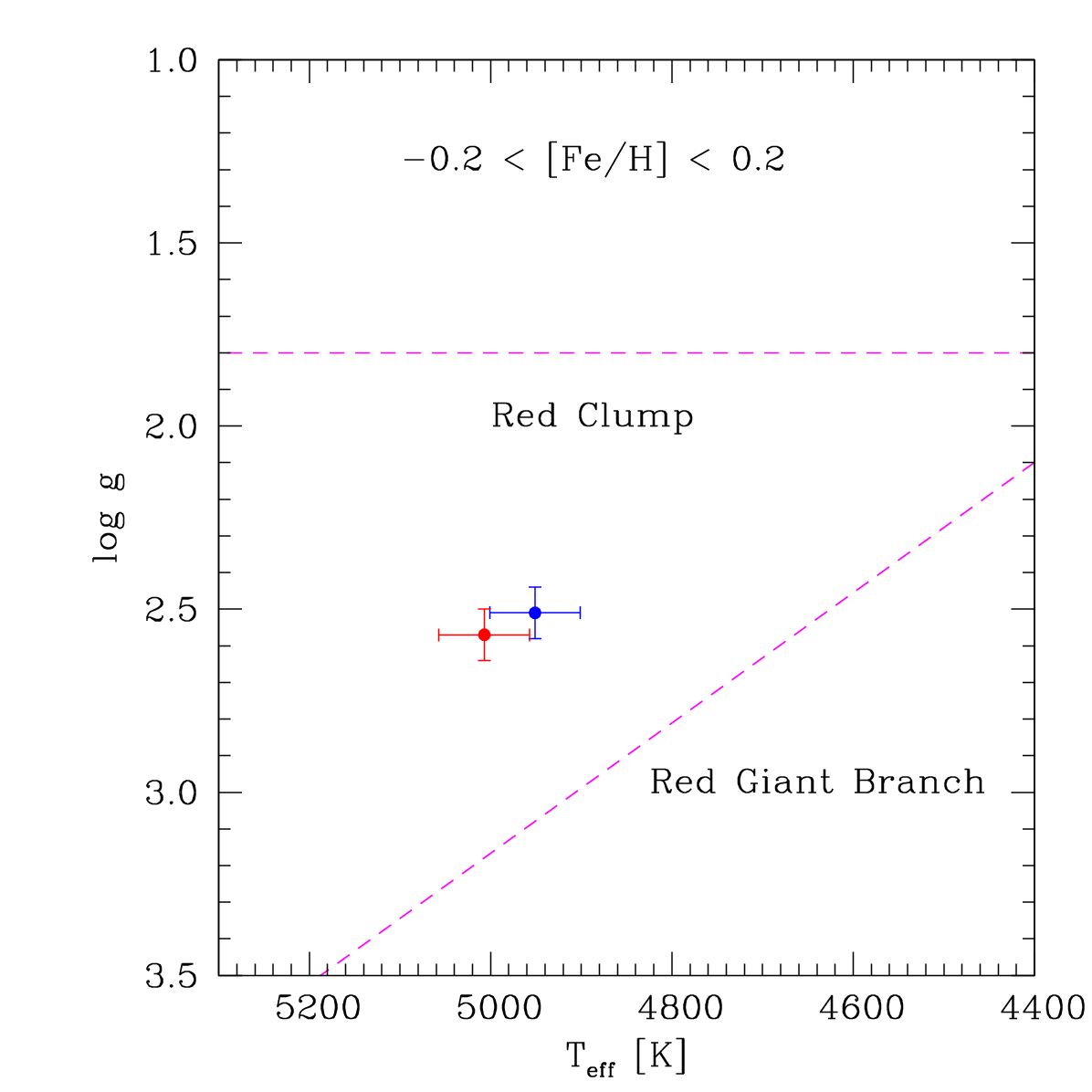}
\caption{Position of stars A and B (red and blue) in the log g - T$_{eff}$ diagram 
and the cuts suggested by \citet{bovy14} (dashed lines) to separate the RGB and RC regions.
The regions correspond to stars in the range -0.2 $<$ [Fe/H] $<$ 0.2.}
\label{Bovy2014.fig}%
\end{figure}

The individual ages shown in Table \ref{mass.age.tab} strongly suggest that both stars belong to the RGB
or that both stars belong to the RC, and are not a mixed pair RGB+RC. 
Although a number of RGB and RC stars overlap in the log T$_{eff}$ $-$ log g diagram, 
\citet{bovy14} suggested a method for distinguishing them. This method requires an accuracy better than 100 K and 0.1 dex in T$_{eff}$ and log g.
We present in Fig. \ref{Bovy2014.fig} the position of stars A and B (red and blue) in the log g - T$_{eff}$ diagram 
and the cuts suggested by \citet{bovy14} (dashed lines) to separate the RGB and RC regions,
which is analogous to Fig. 1 of their work.
Both stars would belong to the RC region in this diagram.
However, we caution that asteroseismic parameters for both stars could help us to distinguish their evolutionary stage better
\citep[e.g. ][]{montalban10,bedding11}.
When we assume that both stars belong to the RGB or RC phase, we can adopt a common (average) age
of 337$\pm$175 Myr or 514$\pm$255 Myr, respectively
(while a mixed pair RGB+RC would present an intermediate age).
The common age for stars A and B presents a considerable uncertainty.
In any case, we show below that chemical evolutionary effects do not explain the observed differences
between stars A and B (for iron and other metals) for any combination of the considered phases.

\begin{figure*}
\centering
\includegraphics[width=8cm]{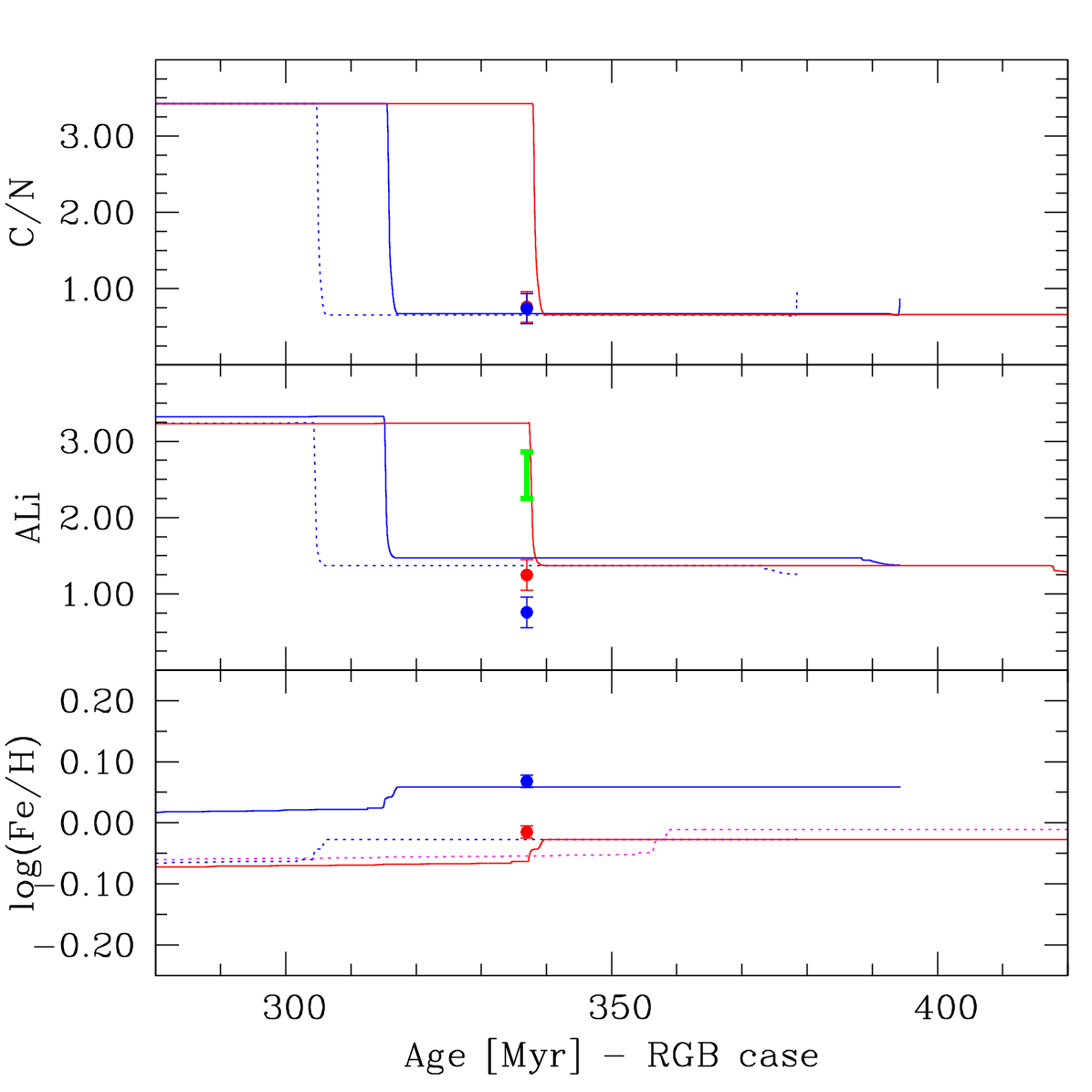}
\includegraphics[width=8cm]{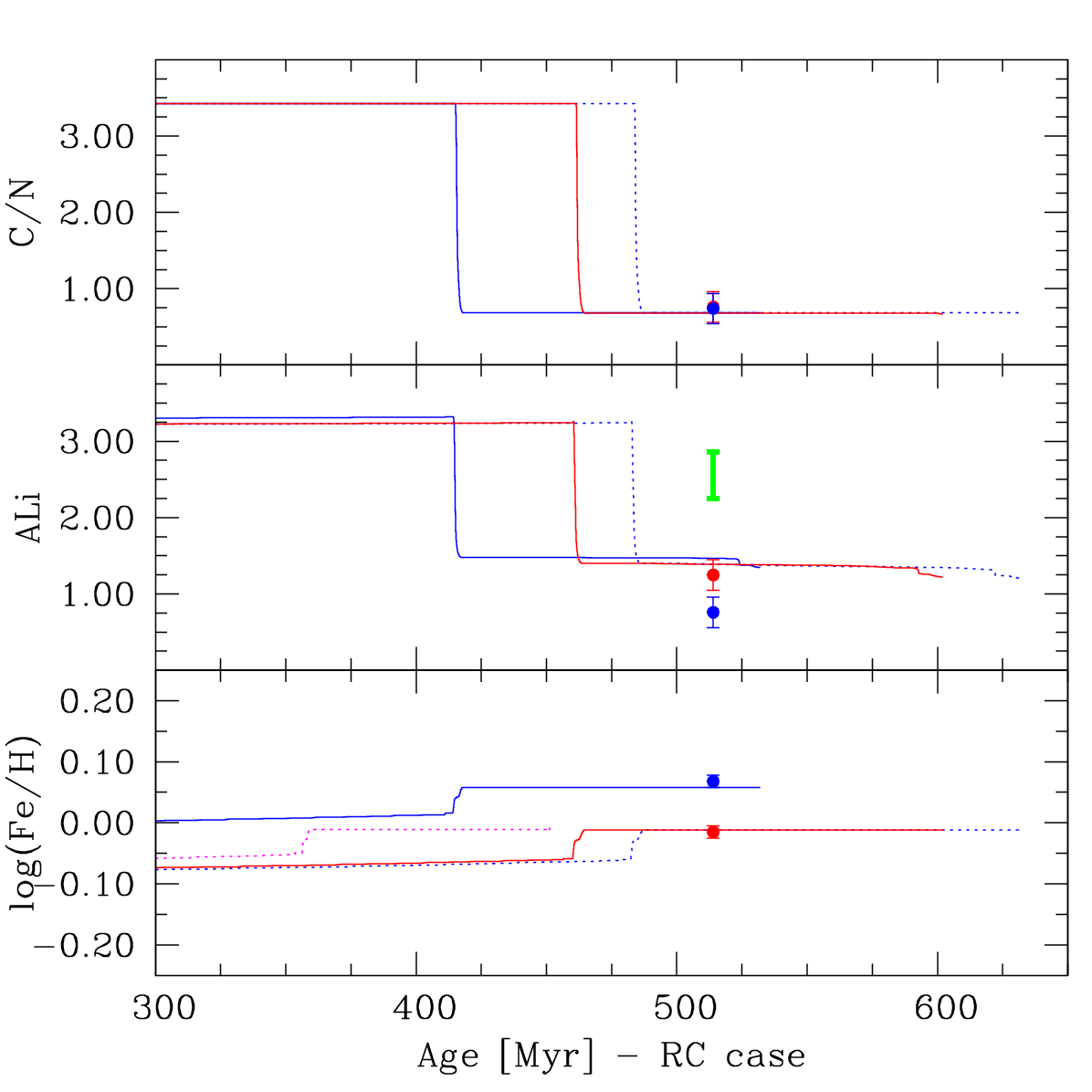}
\caption{C/N, A(Li), and metallicity as a function of age (upper, middle, and lower panels),
estimated with MIST evolutionary tracks (blue and red lines for stars A and B, respectively).
The green bars show the Li content expected in star A after ingestion of planetary material.
The values measured for stars A and B are shown in blue and red (including the error bars on the y-axis).
The left panels correspond to the RGB phase, and the right panels correspond to the RC phase.}
\label{3plots.fig}%
\end{figure*}

During the evolution of giant stars, their extensive envelopes mix internally processed material by fusion
with unprocessed surface material \citep[e.g. ][]{salaris-cassisi17}.
This produces a strong reduction of the superficial C/N ratio and lithium abundances after the FDU
in the RGB phase, for example.
We present in Fig. \ref{3plots.fig} the C/N ratio, the lithium abundance A(Li)\footnote{We define A(Li) $=$ log(Li/H) + 12, similar to \citet{carlberg12}.},
and the metallicity as a function of age, estimated with MIST evolutionary tracks.
The green bars show the Li content expected in star A after ingestion of planetary material (see next sections).
The evolutionary models of stars A and B (blue and red lines) show a significant step or jump in C/N and A(Li),
which corresponds to the strong reduction expected after the FDU.
This step is also present in the evolution of the metallicity, although the jump is considerably lower.
We note that FDU of star A occurs earlier than FDU of star B (the blue step occurs at a younger age than the red step),
which corresponds to the slightly higher mass (and faster evolution) of star A compared to B.
The low values of C/N and A(Li) shown by stars A and B strongly suggest that both stars have already gone through the FDU phase,
(see, e.g., the right panel of Fig. \ref{3plots.fig} for the case of A and B taken as RC stars).
However, for the RGB case of star B (left panel, red), the evolutionary model predicts that the 
FDU will take place in the next few million years, while the low C/N and A(Li) values show that the FDU already occurred in star B.
This disagreement by a few million years shows that the true age of both stars (for the RGB case) might be slightly older
than estimated, or, alternatively, that star B belongs to the RC phase rather than the RGB phase,
as suggested by Fig. \ref{Bovy2014.fig}.

The middle panels of Fig. \ref{3plots.fig} show that the lithium abundances in stars A and B
are lower than the expected values from MIST models.
However, it is difficult to reach a perfect agreement between observations and models because many effects might
modify the lithium abundances.
For instance, the modeled Li content strongly depends on the assumed value for $v/v_{crit}$, which is the initial rotation rate 
(compared to the critical one) used in the MIST models.
We adopted $v/v_{crit}$ $=$ 0.0; but if we had adopted $v/v_{crit}$ $=$ 0.4, the predicted lithium abundances
would be several orders of magnitude lower than observed.
\citet{charbonnel20} showed that the evolution of Li strongly depends on whether rotation is included in the models 
(see, e.g., their Fig. 11).
The evolutionary tracks of MIST also assume a starting value of A(Li) near $\sim$3.3 dex \citep{lodders-fegley98},
while other authors adopted a lower value \citep[3.11 dex, ][]{charbonnel20}.
The number of still uncertain procceses that might modify the lithium abundances led some authors \citep[e.g. ][]{privitera16}
to suggest the use of an increased rotation instead of lithium, in order to detect
a possible planet ingestion (see text above), for example.


The evolutionary effects for the case of iron were estimated using MIST tracks (Fig. \ref{3plots.fig}, lower panels).
The highest variation in [Fe/H] predicted by models during the evolution of stars A and B is $\sim$0.04 dex
(for the RGB or RC cases).
This value corresponds to the increase in metallicity after the FDU, the step in the abundances shown in the Fig. \ref{3plots.fig}.
However, during the evolution after the FDU, the predicted variation in metallicity is almost null (lower than 0.01 dex for both stars).
Thus, we can consider two possiblities:
First, if both stars underwent the FDU before (as strongly suggested by their C/N and Li abundances),
evolutionary effects ($<$0.01 dex) clearly cannot explain the observed difference of $\sim$0.08 dex.
Second, if we assume that only one star passed through the FDU (which is highly unlikely),
evolutionary effects ($<\sim$0.04 dex) are not enough to explain the observed difference of $\sim$0.08 dex.
In this way, evolutionary effects cannot account for the observed differences between stars A and B,
even assuming that only one of the stars underwent the FDU (which is highly unlikely).
This is also valid considering that both stars belong to RGB, both belong to RC, or are a combination of RGB+RC.
Other abundances such as [Si/H], [S/H], [Ca/H], and [Ti/H] were also modeled by MIST and show a step
very similar to the steps of [Fe/H], mainly produced by a slight reduction of the superficial hydrogen
after the FDU \citep{dotter17}.

We also performed additional experiments by calculating hypothetical tracks of star A.
We assumed that star A was born with the same metallicity as star B, but considered different values for its mass:
the same masses previously estimated (Table \ref{mass.age.tab}) for the RGB and RC cases
(dotted blue lines in the Fig. \ref{3plots.fig}), and also a mass of 3 M$_{\odot}$ (dotted magenta line in the lower panel of Fig. \ref{3plots.fig}).
An increase or decrease in the mass of the stars basically corresponds
to a decrease or increase in the age when the FDU occurs (e.g., compare the dotted blue and magenta lines in the lower panel of Fig. \ref{3plots.fig}).
In other words, a modified mass of the stars would basically displace the iron lines mainly along the x-axis in Fig. \ref{3plots.fig}
(on the other hand, a modified starting [Fe/H] would displace the lines mainly along the y-axis).
Therefore, evolutionary effects does not seem to be enough to explain a [Fe/H] difference of $\sim$ 0.08 dex between stars A and B
(both RGB, both RC, or a combination of RGB+RC).

\section{First-order estimation of the ingested mass}

In this section, we estimate the mass ingested by star A, if any, to explain its higher metallicity compared to star B.
We used equation (5) of \citet{teske16} \citep[which is similar to equation (1) of ][]{ramirez11},

\begin{centering}
\begin{equation}
\Delta[M/H] = \mathrm{log}\bigg[\frac{(Z/X)_{cz}\mathcal{M}_{cz} + (Z/X)_P\mathcal{M}_p}{(Z/X)_{cz}(\mathcal{M}_{cz}+\mathcal{M}_p)}\bigg],
\label{eq5}
\end{equation}
\end{centering} 

where $\Delta$[M/H] is the difference in metallicity,
(Z/X)$_{CZ}$ is the ratio of the fractional abundance of metals relative to hydrogen in the convective zone,
M$_{CZ}$ is the mass of the convective envelope,
(Z/X)$_{p}$ is the metallicity of the planet,
and M$_{p}$ is the mass of the planet.
This model assumes an instantaneous accretion and addition of the planet to the convective envelope of the star,
and it should be considered as a first-order approximation.
Adopting from \citet{asplund09} the mass fractions for the Sun, we have
(Z/X)$^{Sun}_{CZ}$ = (0.0134/0.7381) = 0.0182.
We estimated the convection zone metallicity (Z/X)$_{CZ}$
by scaling the solar value with the metallicity \citep[similar to the procedure of ][]{ramirez11,teske16},
(Z/X)$_{CZ}$ $=$ 0.0182 x 10$^{-0.015}$ $\sim$ 0.0176.

In equation \ref{eq5}, we considered two reference values for the planet metallicity,
(Z/X)$_{p}$ $=$ 0.10 and (Z/X)$_{p}$ $=$ 0.36.
The first value of (Z/X)$_{p}$ is similar to that of Jupiter, whose metallicity is estimated to be between 0.04-0.12 \citep{ramirez11}.
For the second value of (Z/X)$_{p}$, we considered an extreme case of metal content inside a planet \citep{thorngren16},
HAT-P-20 b includes 600 M$_{\oplus}$ of metals inside a 7.2 M$_{Jup}$ ($\sim$2290 M$_{\oplus}$) planet.
In this last case, we roughly estimate (Z/X)$_{p}$ $\sim$ 600/(2290-600) $\sim$0.36 for a planet with an extreme metallic content.
However, we note that these massive (and metallic) planets are hard to explain by current planet formation models 
\citep[see e.g. ][]{leconte09,thorngren16},
requiring a migrating planet that accumulates all of the metal available in the disk.

Equation \ref{eq5} also depends on the convective envelope mass M$_{CZ}$ adopted,
which is a function of the stellar age.
Some authors estimated average M$_{CZ}$ values near 77\% of the stellar mass \citep{pasquini07b}
and $\sim$65\% \citep[Figure 1 of ][]{girardi16} for giant stars with masses between 1.0-1.3 M$_{\odot}$.
Other authors assumed that the M$_{CZ}$ encompasses most of the giant star, making fully convective objects,
that is to say, M$_{CZ}$ near 100\% \citep[e.g. ][]{basu-hekker20}.
Thus, for an estimate using reference values, we considered three possible average values for M$_{CZ}$: 
100\%, 77\%, and 35\% of the stellar mass.
In particular, the lower value of M$_{CZ}$ (35\%) would correspond to the most favorable case for the
detection of a metallicity signature of a planetary ingestion.

The next step was to estimate the mass of the planet M$_{p}$ neccesary to explain 
the difference in metallicity observed $\Delta$[M/H]$\sim$0.08 dex.
We estimated M$_{p}$ using equation \ref{eq5}, which depends on the assumed values 
for (Z/X)$_{p}$, M$_{CZ}$, and so on.
As explained, we considered two possible values for (Z/X)$_{p}$ and three possible values for M$_{CZ}$.
The combination of these values produced six different cases with
six different values for M$_{p}$ (upper panel of Table \ref{ingested.mass}).
The resulting values of M$_{p}$ (Col. 4 of Table \ref{ingested.mass})
show their strong dependence on the planet metallicity (Z/X)$_{p}$ and on the
convective envelope mass M$_{CZ}$.
In these six cases, we assumed a mass of 2.872 M$_{\odot}$ for star A,
which would correspond to the RC phase (upper panel of Table \ref{ingested.mass}).
However, if star A belongs to the RGB phase, its mass would be 3.183 M$_{\odot}$,
resulting in six different values for M$_{p}$ (lower panel of Table \ref{ingested.mass}).

In addition, we took the opportunity and estimated the expected increase in the
rotational velocity after the planet ingestion,
$\Delta$v$_{rot}$ $=$ v$_{rot}$(final) $-$ v$_{rot}$(initial),
using equation (1) of \citet{carlberg09}.
We also estimated the expected increase in the lithium abundance
after the planet ingestion, $\Delta$Li $=$ A(Li)$_{final}$ $-$ A(Li)$_{initial}$,
using equation (2) of \citet{carlberg12}.
We adopted A(Li)$_{initial}=$-0.18$\pm$0.08 dex similar to \citet{carlberg12}, which is the average abundance of giant slow rotators.
Both $\Delta$v$_{rot}$ and $\Delta$Li depend on M$_{p}$ and M$_{CZ}$,
and they are shown in the two last columns of Table \ref{ingested.mass}.
In order to estimate $\Delta$v$_{rot}$, 
we adopted a hot Jupiter-like initial value of the semimajor axis (a$=$0.02 au) for the ingested planet.
(For comparison, a warm Jupiter-like planet with $\sim$0.5 au would imply
a much higher reservoir of angular momentum, resulting in a significantly higher $\Delta$v$_{rot}$
by several orders of magnitude).
Finally, in the estimation of $\Delta$v$_{rot}$ we adopted a null eccentricity (e$=$0),
as assumed by \citet{carlberg09} for planets with unknown eccentricity.
In this way, Table \ref{ingested.mass} summarizes the effects after a planet ingestion,
including an estimate of the required mass to ingest M$_{p}$,
together with the expected $\Delta$v$_{rot}$ and $\Delta$Li values after the pollution event.

\begin{table*}
\centering
\caption{Estimate of the required planetary mass (M$_{p}$), 
variation in rotation ($\Delta$v$_{rot}$), and lithium content ($\Delta$Li)
after a possible ingestion event.}
\begin{tabular}{cccccc}
\hline
\hline
Case & M$_{CZ}$ & (Z/X)$_{p}$ & M$_{p}$ & $\Delta$v$_{rot}$ & $\Delta$Li \\
     &          &             & [M$_{Jup}$] & [km s$^{-1}$] & [dex] \\
\hline
Star A in the RC phase: \\
\hline
 1 & 100\% & 0.10 & 135.0  & 37.46 & +2.10 \\
 2 & 100\% & 0.36 & 31.5   & 8.74  & +1.49 \\
 3 & 77\%  & 0.10 & 105.0  & 37.84 & +2.10 \\
 4 & 77\%  & 0.36 & 24.2   & 8.72  & +1.49 \\
 5 & 35\%  & 0.10 & 47.5   & 37.66 & +2.10 \\
 6 & 35\%  & 0.36 & 11.0   & 8.72  & +1.49 \\
\hline
Star A in the RGB phase: \\
\hline
 1 & 100\% & 0.10 & 150.0  & 37.88 & +2.10 \\
 2 & 100\% & 0.36 & 35.0   & 8.84  & +1.49 \\
 3 & 77\%  & 0.10 & 115.0  & 37.71 & +2.10 \\
 4 & 77\%  & 0.36 & 27.0   & 8.85  & +1.49 \\
 5 & 35\%  & 0.10 & 52.5   & 37.88 & +2.10 \\
 6 & 35\%  & 0.36 & 12.2   & 8.80  & +1.49 \\
\hline
\end{tabular}
\normalsize
\label{ingested.mass}
\end{table*}

\begin{figure*}
\centering
\includegraphics[width=8cm]{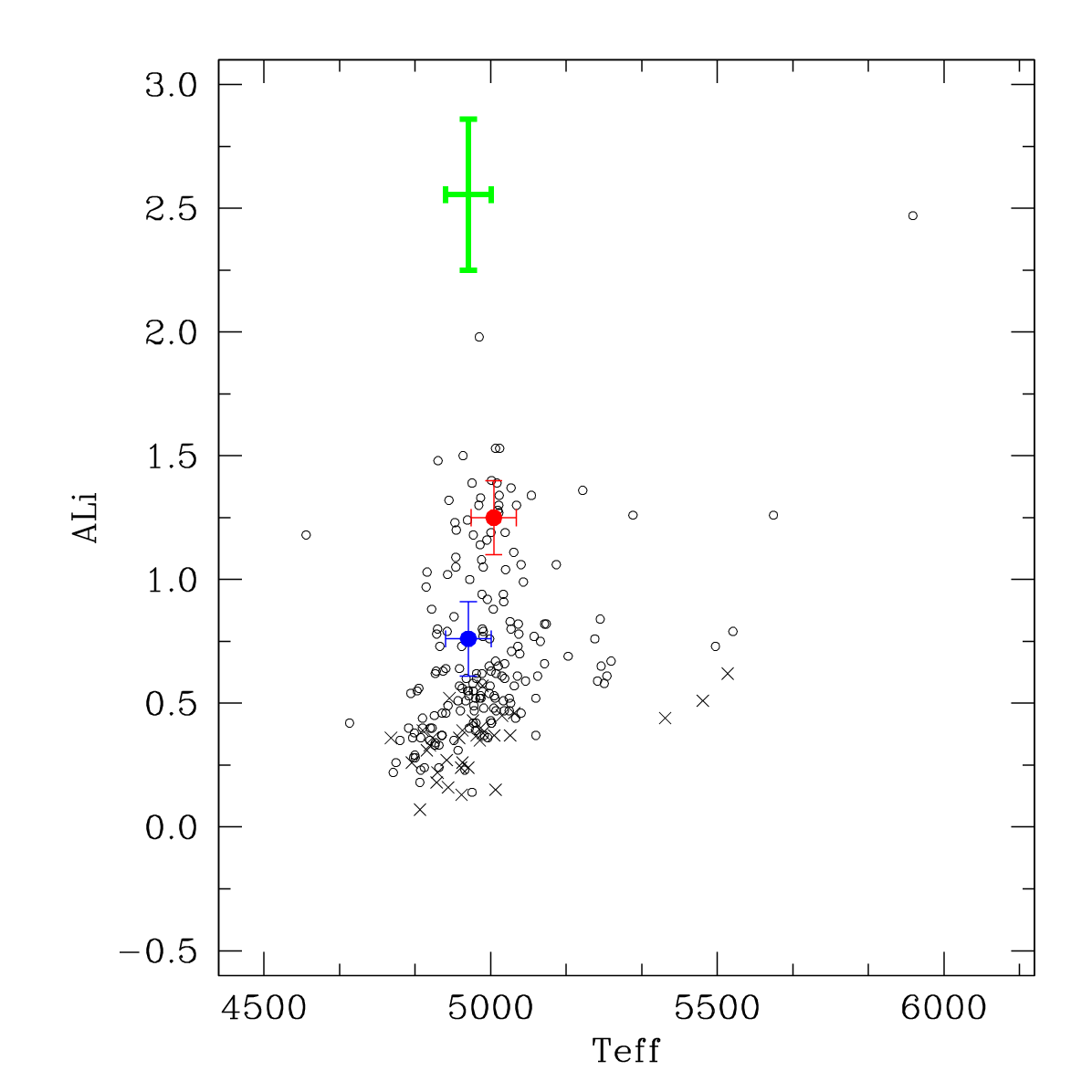}
\includegraphics[width=8cm]{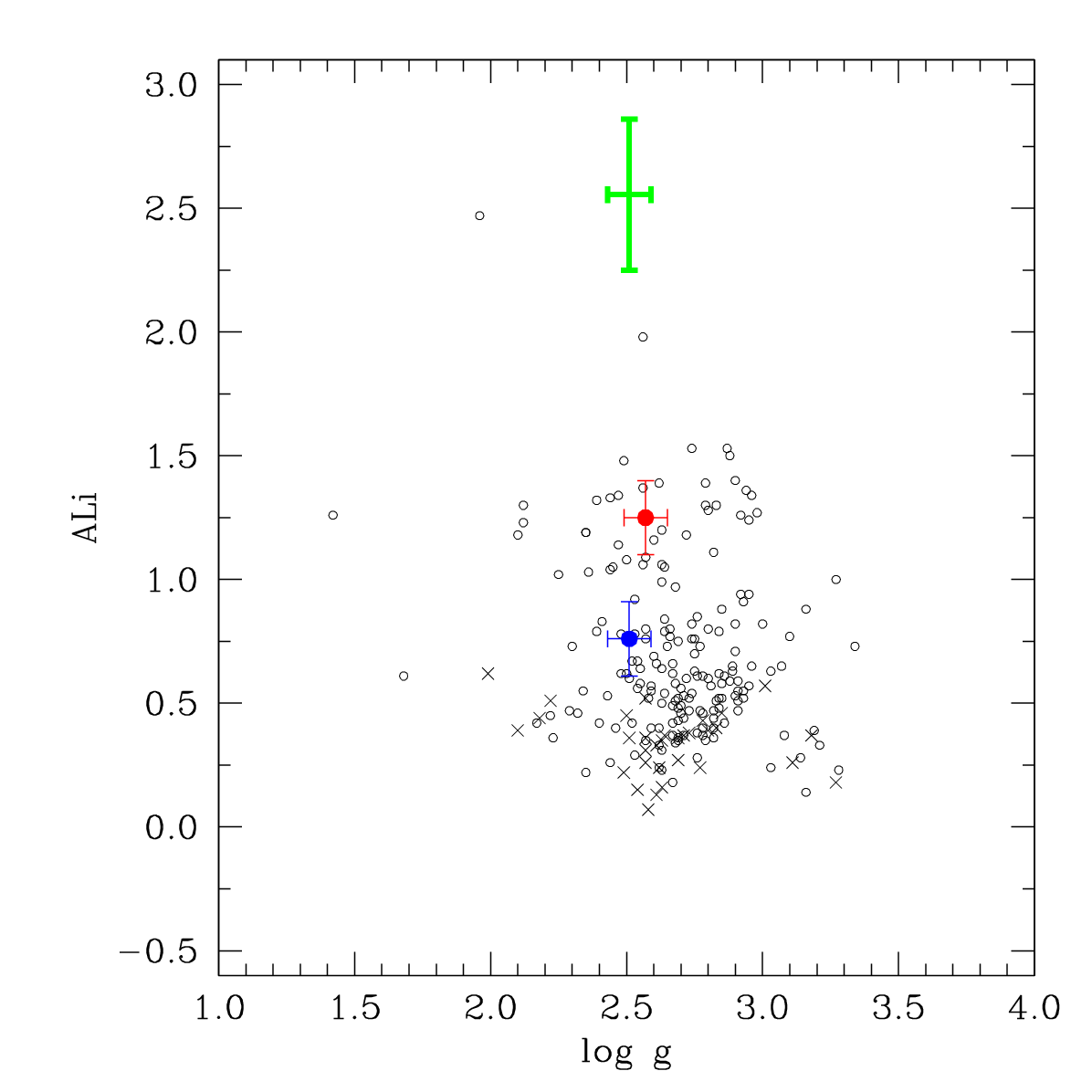}
\caption{Li abundance derived for 378 G/K giant stars
from \citet{otro.liu14} as a function of T$_{eff}$ and log g (left and right panels).
The empty circles and crosses correspond to Li values and upper limits, respectively, measured for the 378 giant stars.
The measured values in this work for stars A and B are shown in blue and red, respectively,
and the predicted values after a possible planetary ingestion in star A are shown in green.}
\label{li.vs.params.fig}%
\end{figure*}

In order to study a possible planetary ingestion, we also compared the lithium
content in stars A and B with that of other giant stars.
We present in Fig. \ref{li.vs.params.fig} the Li abundance derived for 378 G/K giant stars
from \citet{otro.liu14} as a function of T$_{eff}$ and log g (left and right panels).
The empty circles and crosses correspond to Li values and upper limits, respectively, measured for the 378 giant stars.
We restricted the stars shown to a metallicity range between -0.2 dex and 0.1 dex, 
which is closer to the values of the stars in this work.
The measured values in this work for stars A and B are shown in blue and red, respectively,
and the predicted values after a possible planetary ingestion in star A are shown in green.
The green lines correspond to possible $\Delta$Li values between 1.49 dex and 2.10 dex estimated for star A
(Table \ref{ingested.mass}).
Li abundances of \citet{otro.liu14} were corrected for NLTE effects by interpolating in 
the data of \citet{lind09}. In order to properly compare the abundances,
we also corrected for NLTE following the corrections of \citet{lind09}. Fig. \ref{li.vs.params.fig} shows a considerable dispersion of Li content for a fixed value of the parameters.
The Li abundances of stars A and B appear to be similar to those of other giant stars.
\citet{otro.liu14} suggested a trend in which giant stars with higher T$_{eff}$ tend to show a higher lithium content.
The lower Li in star A compared to star B might at least partially be explained by the higher T$_{eff}$ of star B.
On the other hand, a planetary ingestion event would produce a significant increase in the Li content in star A
(the green values shown in Fig. \ref{li.vs.params.fig}),
being considerably higher than current A(Li) observed in both stars A and B.
However, Fig. \ref{li.vs.params.fig} shows no evidence of an increased Li content in star A
(on the contrary, star A displays a lower Li content than star B).
In other words, there is no need to invoke an ingestion event to explain the Li content in stars A and B.

We also considered the possibility of a brown dwarf (BD) ingestion by star A.
We focused on the most favorable conditions for the detection of a chemical difference,
that is, by taking M$_{CZ}$ $\sim$ 35\% and star A in the RC phase (mass of 2.87 M$_{\odot}$).
For the metallicity of the ingested object (Z/X)$_{p}$ (in this case, a brown dwarf, (Z/X)$_{BD}$), we also scaled the solar value
by the metallicity, that is, (Z/X)$_{BD}$ $=$ 0.0182 x 10$^{+0.3}$ $\sim$ 0.0363.
The metallicities of known BDs range between -1.0 to +0.3 dex \citep[e.g. ][ and references therein]{meisner23},
and we adopted a metal-rich BD with [Fe/H]=+0.3 dex.
In this case, equation \ref{eq5} implies that it would be necessary to ingest more than $\sim$250 M$_{Jup}$
of material to reach a difference of $\sim$0.08 dex between stars A and B. This is beyond the mass limits of BDs. 
These high mass values would also imply a significant increase in the rotational velocities
and lithium content, which is not detected in the binary system of this work.

The question now is when the possible ingestion occurred.
A number of events might modify the superficial abundances of giant stars, such as the FDU,
and therefore, it is relevant to consider this question.
However, it is difficult to give a precise estimate.
Equation \ref{eq5} used in this work to estimate the possible amount of ingested mass
assumes an instantaneous engulfment, that is, a very recent event.
Similarly, equation (1) of \citet{carlberg09} (which describes the angular momentum deposition of the planet)
and equation (2) of \citet{carlberg12} (which estimates the increase in lithium after the planet ingestion)
were derived without specific time or evolutionary constraints.
However, we can consider the following.
The values of the C/N ratio and A(Li) strongly suggest that the giant stars A and B already underwent the FDU.
The values shown in the Table \ref{ingested.mass} show that 
it would be easy to detect a planet ingestion for a lower convective envelope mass ($\sim$35\%) than for a massive convective envelope ($\sim$100\%).
This suggests that a planet ingestion event would be more easily detected after the FDU rather than
before the FDU.
Similarly, \citet{aguilera-gomez16} suggested that ingestion events would be easily detected after the FDU and
before the RGB bump.

For a moment, we assume the following scenario: The planet ingestion occurred after the FDU in star A, 
and several million years after this event, the convective envelope mixed and erased the additional lithium that was added by the ingestion
\citep[this lithium depletion is not considered by equation (2) of ][]{carlberg12},
perhaps bringing the lithium abundances to the values currently observed in star A.
However, in this case, it would be difficult to explain why the same mixing
(which easily restored the lithium abundance to pre-ingestion values)
was unable to restore other metals to their pre-ingestion values.
In other words, the massive convective envelope should diminish the additional lithium added by the ingestion
(by mixing with layers with a a much lower lithium content, and perhaps also by reaching
deeper and hotter regions in which lithium could be destroyed). That is, to diminish the additional lithium,
significant mixing is required. However, in this case, it would not be clear why the mixing
(which was significant for lithium) was not significant for other metals as well.
Even assuming that the mixing erased the ingested lithium but not other metals
(which is highly unlikely),
the values of Table \ref{ingested.mass} for the case of convective envelopes of $\sim$35\% 
suggest that we would require the ingestion of massive planets ($>$ 11 M$_{Jup}$) with
an extremely high metallic content.
As mentioned previously, these extremely metallic planets are very difficult to explain with current planet formation models
and would require the accretion of almost all metals available in the disk at the time of planet formation.
Similarly, in this case, we still have to explain why there is no evidence of an increased rotational velocity
in star A compared to star B, and currently, no planets are detected orbiting stars A or B.
Formally speaking, although previous arguments do not completely rule out an ingestion event in star A,
we consider that an engulfment event is highly unlikely.

\end{appendix}

\end{document}